\documentclass[%
 reprint,
 amsmath,amssymb,
 aps,
]{revtex4-2}

\usepackage{graphicx}
\usepackage{dcolumn}
\usepackage{bm}

\usepackage[caption=false]{subfig}
\usepackage{xcolor}
\usepackage{centernot}
\usepackage{braket}

\begin{document}

\preprint{APS/123-QED}

\title{Union--find quantum decoding without union--find}

\author{Sam J.\ Griffiths}
\author{Dan E.\ Browne}%
\affiliation{%
 Dept.\ of Physics and Astronomy, University College London, London, WC1E 6BT, United Kingdom
}%


\date{\today}

\begin{abstract}
The union--find decoder is a leading algorithmic approach to the correction of quantum errors on the surface code, achieving code thresholds comparable to minimum-weight perfect matching (MWPM) with amortised computational time scaling near-linearly in the number of physical qubits. This complexity is achieved via optimisations provided by the disjoint-set data structure. We demonstrate, however, that the behaviour of the decoder at scale underutilises this data structure for twofold analytic and algorithmic reasons, and that improvements and simplifications can be made to architectural designs to reduce resource overhead in practice. To reinforce this, we model the behaviour of erasure clusters formed by the decoder and show that there does not exist a percolation threshold within the data structure for any mode of operation. This yields a linear-time worst-case complexity for the decoder at scale, even with a naive implementation omitting popular optimisations.
\end{abstract}

\maketitle


\section{\label{sec:introduction}Introduction}

Quantum error correction (QEC) is considered essential for the development of scalable, fault-tolerant quantum computers in the medium- to long-term \cite{nielsenQuantumComputationQuantum2002}. One of the most dominant approaches in the field of QEC is topological error correction, commonly referred to as \emph{surface codes} \cite{dennisTopologicalQuantumMemory2002, fowlerSurfaceCodesPractical2012}. In such codes, qubits are arranged on the surface of some topology, transforming the problem of QEC into one of (largely classical) graph theory and algorithm design. By taking parity checks, a \emph{syndrome} can be obtained, giving information on stochastic errors without decohering the computational states.

An optimal decoder, inferring a correction operator from a syndrome with maximum success likelihood, is computationally hard, scaling exponentially in the number of physical qubits. Minimum-weight perfect matching (MWPM) instead returns the single most likely error configuration in polynomial time \cite{dennisTopologicalQuantumMemory2002}, although recent work has improved this to near-linear time in sparse regimes \cite{higgottSparseBlossomCorrecting2023}.

Another leading approach is the \emph{union--find decoder} \cite{delfosseLineartimeMaximumLikelihood2017, delfosseAlmostlinearTimeDecoding2017}, which instead uses a cluster-growing method to find results very similar to MWPM in effectively linear time. Proposals for the architectural implementation of the decoder, complete with analysis of resource consumption, have been made, such as the AFS architecture \cite{dasAFSAccurateFast2022}. The union--find decoder is so named after reliance on the disjoint-set data structure providing its enviable time complexity via some key algorithmic optimisations \cite{tarjanWorstcaseAnalysisSet1984}. 

In this work, the behaviour of the union--find decoder and its usage of the disjoint-set data structure is studied. In particular, we reason that the data structure is underutilised and that the implementation can thus be simplified, potentially saving both time and memory in what is widely considered a critical bottleneck for fault-tolerant quantum computing. We present results and analysis regarding the behaviour of the union--find decoder at scale in Section~\ref{sec:results} by first studying the complexity of the disjoint-set data structure, then gathering data about the performance of the decoder, then finally linking these topics analytically with a percolation model.

We use the AFS architecture as a framework for simulating the union--find decoder; discussion of our implementation and improvements can be found in Appendix~\ref{appendix:afs}.

\section{\label{sec:background}Background}
\subsection{Topological QEC}
Classical error correction is generally based upon introducing redundancy to protect information, by encoding logical bits into a subspace of a greater number of physical bits \cite{macwilliamsTheoryErrorcorrectingCodes1977}. A \emph{repetition code} is the most basic example of this type of scheme. For example, $n=3$ physical bits can be used to encode one logical bit with the mapping
\begin{align}
    0_L = 000, 1_L = 111
\end{align}
where 000 and 111 are known as the $codewords$. If the measured state is outside of the codespace (i.e.\ it is neither 000 nor 111) then at least one bit-flip error must have occurred. Error correction can be implemented by simply taking a majority vote. If each physical bit is subject to an independent error rate $p$, then the logical error rate of this scheme (that is, the probability of the majority vote being incorrect, which happens if over half of the bits were flipped) becomes
\begin{equation}
    p_L = p^3 + 2p^2(1-p)
\end{equation}
$p_L < p$ if $p < 0.5$, which means that the scheme successfully reduces the error rate as long as $p < 0.5$, which is known as the \emph{code threshold} $p_c$. Sub-threshold, increasing $n$ suppresses $p_L$ arbitrarily close to zero. Therefore, the code threshold is a measure of the fault-tolerance of a given scheme at scale.

In quantum information, bit-flips (Pauli $X$ noise) can be corrected using an equivalent encoding for logical qubits, e.g.\
\begin{align}
    \ket{0_L} &= \ket{0}\otimes\ket{0}\otimes\ket{0} = \ket{000} \\
    \ket{1_L} &= \ket{1}\otimes\ket{1}\otimes\ket{1} = \ket{111}
\end{align}
The computational basis cannot be measured without decoherence, so parity checks can instead be taken \cite{nielsenQuantumComputationQuantum2002}. In this example, the observables $ZZI$ and $IZZ$ would reveal where neighbouring qubits differ in parity. These parity measurements are known as the $error syndrome$. In effect, this implements the majority vote scheme without decoherence.

Phase-flip errors (Pauli $Z$ noise) can equivalently be detected with the encoding
\begin{align}
    \ket{0_L} &= \ket{+++} = (\ket{000}+\ket{111})/\sqrt{2} \\
    \ket{1_L} &= \ket{---} = (\ket{000}-\ket{111})/\sqrt{2}
\end{align}
and the parity checks $XXI$ and $IXX$. As any arbitrary error operator $E$ can be decomposed in the form
\begin{equation}
    E = e_0I + e_1X + e_2Z + e_3XZ
\end{equation}
it suffices to correct $X$ and $Z$ noise independently in order to correct for arbitrary errors. This technique is often assumed when discussing QEC codes. Nesting the above $n=3$ encodings using $n=9$ thus corrects for any error on a single physical qubit. Codes with this property of independently correcting $X$ and $Z$ noise are known as \emph{CSS codes} \cite{calderbankGoodQuantumErrorcorrecting1996, steaneMultipleparticleInterferenceQuantum1997}.

\begin{figure}
    \centering
    \includegraphics[width=\linewidth]{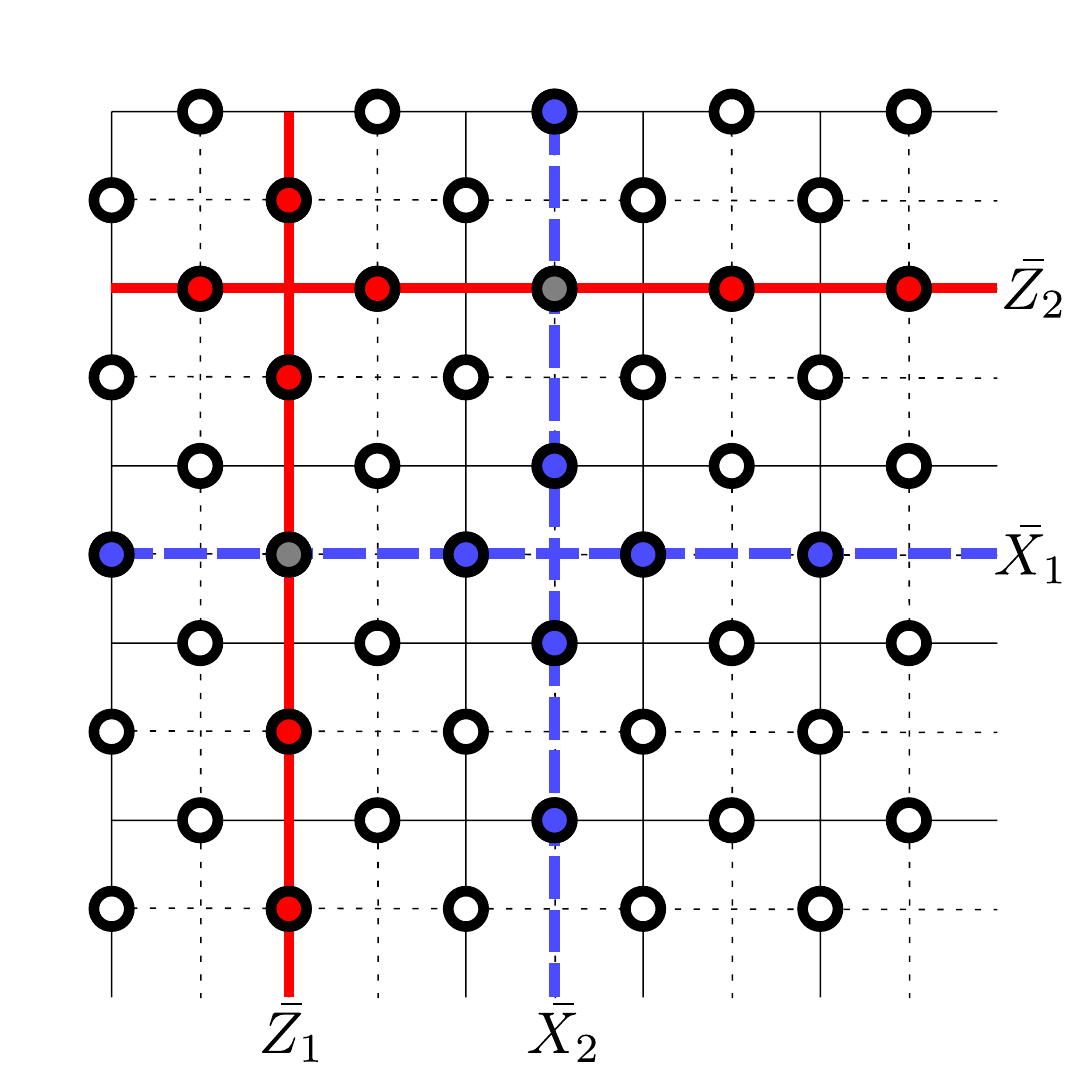}
    \caption{Toric code with $L=5$. Physical qubits are circles on the edges of the latice. The dashed lines show the dual lattice. Logical operators $Z$ and $X$ for two logical qubits are shown as nontrivial cycles on the primal and dual lattices, respectively.}
    \label{fig:toric-code}
\end{figure}

In general, more sophisticated codes can encode multiple logical qubits and protect against more than one physical error. If the repetition code is viewed as a 1-dimensional string, this generalises to physical qubits on $n$-dimensional topologies. The motivating example of topological QEC is the \emph{toric code} \cite{kitaevFaulttolerantQuantumComputation1997, dennisTopologicalQuantumMemory2002}. Parity checks are arranged as vertices on a 2-dimensional lattice with periodic boundary conditions, as on the surface of a torus (Fig.~\ref{fig:toric-code}). The edges between these vertices are physical qubits. A plaquette operator is defined as the combination of four $Z$ operators on neighbouring qubits, forming a trivial cycle on the primal lattice. Similarly, a vertex operator is the combination of four $X$ operators forming a trivial cycle on the dual lattice.

A lattice of length $L$ has $n=2L^2$ physical qubits, $L^2$ plaquette operators and $L^2$ vertex operators, such that the number of logical qubits is $k=2L^2-2(L^2-1)=2$. Logical $Z$ operators are defined as nontrivial cycles on the primal lattice and logical $X$ operators as nontrivial cycles on the dual lattice (Fig.~\ref{fig:toric-code}).

A parity check yields $-1$ if it is incident to an odd number of Pauli errors. Therefore, the error syndrome acts to flag endpoints of strings of neighbouring Pauli errors. The single most likely error configuration, assuming an independent error model, is the subgraph which pairs excited syndrome vertices with the fewest edges. This problem is known formally as the \emph{minimum-weight perfect matching} (MWPM). The MWPM decoder takes this subgraph as the correction operator, achieving a code threshold of ${\sim}10.3\%$, not much below the code threshold of the optimal decoder at ${\sim}11.0\%$. The code threshold is the value of the independent error rate $p$ below which the logical error rate is suppressed to zero with lattice size $L \to \infty$ \cite{dennisTopologicalQuantumMemory2002, calderbankGoodQuantumErrorcorrecting1996, wangConfinementHiggsTransitionDisordered2003}.

Note that whilst the size of the lattice is often given as the length $L$, the \emph{code distance} $d$ is also commonly used, which can be defined as the minimum weight of an undetectable error. In the case of the $L \times L$ toric code, $d=L$, but this is not exactly true in other cases, such as the \emph{planar code} (with hard boundary conditions instead of periodic) \cite{dennisTopologicalQuantumMemory2002}. In this paper, we may refer to $L$ and $d$ interchangeably when representing lattice size, depending on the specific context.

\subsection{\label{subsec:union--find}Union--find decoder}
The best-known algorithms for MWPM are the \emph{blossom algorithm} and its derivatives, with runtime scaling polynomially in the number of qubits $n$ \cite{edmondsPathsTreesFlowers1965, kolmogorovBlossomNewImplementation2009}. This forms a popular benchmark against which alternative decoders are developed, attempting to improve time complexity whilst retaining relatively high code threshold. For example, the union--find decoder runs in near-linear time whilst still achieving ${\sim}9.9\%$ code threshold \cite{delfosseAlmostlinearTimeDecoding2017}. A full description of the decoder can be found in \cite{delfosseLineartimeMaximumLikelihood2017} and \cite{delfosseAlmostlinearTimeDecoding2017}, but to summarise, the decoder comprises two stages: first, \emph{syndrome validation} is the act of converting the Pauli error syndrome into an erasure. Erasure errors represent processes such as qubit loss or leakage and are equivalent to random Pauli errors with positional information; this additional information makes them theoretically easier to decode than stochastic Pauli errors \cite{delfosseLineartimeMaximumLikelihood2017}.

Syndrome validation is effectively a clustering algorithm: clusters are iteratively grown from each syndrome vertex, merging with each other until they support an even number of syndrome vertices, at which point they cease growing. The clusters are disjoint sets of vertices and so can be tracked using the disjoint-set, a.k.a.\ \emph{union--find}, data structure (see Section~\ref{subsec:disjoint-set}). This provides a time complexity of $O(n\alpha(n))$, where $\alpha(n)$ is the extremely slow-growing inverse Ackermann function, such that the complexity is effectively linear in $n$ for all realistic inputs \cite{tarjanWorstcaseAnalysisSet1984}.

The second stage of the decoder is the use of the \emph{peeling decoder} to decode the erasure. First, cycles are removed (trivially, in $O(n)$, via depth-first search or similar) to obtain a spanning forest on the erasure, which guarantees the time complexity hereinafter. Then, leaf edges are iteratively pruned until the only edges remaining form a maximum-likelihood cover of the original syndrome, using simple decision rules \cite{delfosseLineartimeMaximumLikelihood2017}. This peeling algorithm takes $O(n)$ time, giving an overall complexity of $O(n\alpha(n))$ for the union--find decoder.

The rigorous theoretical justification for the decoder is the efficient conversion of Pauli errors (plus any naturally-occurring erasure errors) to exclusively an erasure, which can be decoded with maximum likelihood in linear time. Practically speaking, however, the decoder can instead be understood as a method to approximate MWPM graphically by growing and then pruning clusters. Its main benefit over MWPM is that it is far more straightforward to implement in a limited hardware regime than heavily-optimised blossom algorithms, whilst also achieving superior time complexity.

Variants of the union--find decoder for weighted graphs, which can incorporate more information of noise biases, have been proposed \cite{huangFaulttolerantWeightedUnionfind2020, wuInterpretationUnionfindDecoder2022}, as have generalisations to quantum LDPC codes \cite{delfosseUnionfindDecoderQuantum2022}. Preprocessing decoding problems with machine learning techniques has also shown promise \cite{meinerzScalableNeuralDecoder2022}. A distributed version of the decoder has also been recently proposed and demonstrated on FPGA hardware \cite{liyanageScalableQuantumError2023} and extended with a strictly-local version \cite{chanStrictlyLocalUnionfind2023}.

Finally, note that the above code thresholds have assumed the use of 2D lattices with an independent qubit error rate $p$. This approach remains susceptible to errors in the measurement of syndrome checks. A popular solution is to take repeated rounds of measurements, which can be stacked graphically as a 3D problem. Taking the difference syndrome, wherein each vertex is switched on if its value differs from the previous round, allows measurement errors to manifest as vertical timelike errors. The decoder (MWPM, union--find or otherwise) can then often merely be generalised to 3D. This is represented by the \emph{phenomenological error model}, with measurement error rate $q$ distinct to $p$. Code thresholds are generally an order of magnitude lower in this regime \cite{dennisTopologicalQuantumMemory2002, skoricParallelWindowDecoding2023}. 

\subsection{\label{subsec:afs}AFS decoder}
As the decoding layer forms a key bottleneck on the quantum computing stack, decoders with superior time complexity, such as union--find, are valued highly. Memory is also likely to be a restricted resource, not just in terms of total overhead but particularly when considering that the hardware is likely to be situated in a cryogenic environment, limiting the memory available. The AFS (\emph{Accurate, Fast and Scalable}) decoder proposes a design for a computational architecture implementing the union--find decoder \cite{dasAFSAccurateFast2022}.

The architecture comprises three pipelined modules: the \emph{graph generator}, \emph{DFS engine} and \emph{correction engine}, which implement syndrome validation, cycle removal and the peeling decoder, respectively.

The graph generator is the most significant component of the pipeline in terms of both time and memory requirements. A \emph{spanning tree memory} (STM) stores the growth state of the erasure on the lattice, with one bit per vertex and two bits per edge (so that clusters can grow by one half-edge per iteration, such that adjacent clusters merge in one iteration).

The DFS and correction engines are notably simpler in comparison and account for significantly less workload and memory. As the remainder of our paper is primarily concerned with syndrome validation, details of these other modules are left as further reading. The AFS paper also includes models of time and memory requirements, considerations of the distribution of concurrent modules for online error correction at scale, and methods of compressing syndrome data to work within limited bandwidth \cite{dasAFSAccurateFast2022}.

In this work, simulations of both toric and planar codes were developed with MWPM and union--find decoders, as well as a simulation of the AFS architecture on planar codes specifically. The performance results in Section~\ref{subsec:redundancy} are from simulations of the AFS architecture and thus use planar codes, whilst the erasure percolation results in Section~\ref{subsec:percolation} use toric codes simply for ease of definition. Understanding of the AFS architecture specifically is, therefore, irrelevant to our main results and conclusions, although a summary of the improvements we made in its implementation can be found in Appendix~\ref{appendix:afs}.

\subsection{\label{subsec:disjoint-set}Disjoint-set data structure}
Proposals of the union--find decoder have been heavily motivated by the complexity behaviour provided by the underlying disjoint-set, a.k.a.\ union--find, data structure \cite{delfosseLineartimeMaximumLikelihood2017, delfosseAlmostlinearTimeDecoding2017}. This data structure is an efficient method for tracking disjoint, i.e.\ non-overlapping, sets of elements \cite{tarjanWorstcaseAnalysisSet1984}. Elements are stored as a forest, initialised to singleton nodes with no connecting edges. The \emph{union} operation merges two sets into one. For example, if two singleton elements are merged, an edge is added between them. The graph is directional, so trees are formed: one of the elements becomes the parent of the other at random.

The \emph{find} operation, given any single element, returns the root of the tree it belongs to. The roots of the trees are taken as the characteristic elements of the sets. Therefore, querying if any two elements belong to the same set is achieved by checking if the roots returned by the \emph{find} operation on each of them are equal. When performing a \emph{union} operation, the \emph{find} operation is first performed on each in order to find their respective roots, with one of those roots becoming the parent of the other.

In this primitive form, the only required component is a table of parent pointers, hereafter referred to as a \emph{root table}. The time taken to scale a tree in the \emph{find} operation in the average/worst case is proportional to the height of the tree. As it stands, the worst-case scenario of arbitrary tree structure sees the trees degenerate to linked lists, such that the average/worst-case complexity of the \emph{find} operation is $O(n)$. Performing $m$ merge operations would therefore have complexity $O(mn)$, giving the union--find decoder a complexity of $O(n^2)$.

Two optimisations are commonly used to improve this complexity. Firstly, \emph{union-by-size} (UBS) records the size of each set, i.e.\ the number of elements in each tree. When merging two sets, instead of selecting which root becomes the merged root at random, the root of the larger tree is always selected, such that the smaller set is always merged into the larger set. This bounds the heights of trees to $O(\log n)$, so performing $m$ merge operations would have complexity $O(m\log n)$, giving the decoder a complexity of $O(n\log n)$.

As a downside, UBS requires storing a size value for each element in addition to a parent pointer. In this setting, a root table stores the parent pointers in the disjoint set forest (with one index/pointer per vertex) and a size stores the set sizes for UBS (with one integer per vertex). This can be seen, for example, in the AFS architecture.

Secondly, \emph{path compression} tracks all of the elements visited during a \emph{find} operation and, upon finding the root, sets the parents of all visited elements to the root. This flattens the height of each of these elements to 1 for subsequent operations. Nontrivial analysis shows that $m$ merge operations with both UBS and path compression has complexity $O(m\alpha(n))$, giving an amortised complexity per operation of $O(\alpha(n))$ and the decoder its stated complexity of $O(n\alpha(n))$. $\alpha(n)$ is the inverse Ackermann function, which is so slow-growing that the amortised complexity per operation is effectively constant-time and the decoder is overall effectively linear-time.

As a downside, path compression requires storing and performing a second pass over the visited elements. For example, in the AFS architecture, \emph{tree traversal registers} are allocated in order to track visited elements for path compression. To avoid this, \emph{path splitting} instead sets the parent pointer for each element to its grandparent during the single pass, which achieves the same complexity asymptotically. These complexities can be readily observed from a Monte Carlo simulation in Figs~\ref{subfig:ds-per-m-naive}, \ref{subfig:ds-per-m-ubs} and \ref{subfig:ds-per-m-ubs-compression}.

\subsection{Motivation and contributions}
The performance of a decoder algorithm is crucial in a QEC pipeline; classical in nature, it can form a critical bottleneck on the overall performance of an error-corrected quantum computer. This is a key reason for the popularity of the union--find decoder, with its near-linear time complexity offering a significant improvement over MWPM's polynomial complexity.

In this paper, our contributions are in studying the behaviour of the union--find decoding algorithm at scale. In Section~\ref{subsec:saturation}, we show that the complexities of the disjoint-set data structure described above can deviate from the theoretical trend; that is, we demonstrate the existence of an \emph{unsaturated regime} within which the structure is underutilised and the complexity decreases. In Section~\ref{subsec:redundancy}, we describe empirical results which suggest that including the optimisations of UBS and path compression -- commonly assumed to be beneficial -- may offer no improvement in runtime and in fact add overhead overall. In Section~\ref{subsec:percolation}, we define an analytical model of \emph{erasure percolation}, describing the workload placed upon the data structure at scale and showing that a decoder runs strictly in the unsaturated regime. We conclude that the union--find decoder has a complexity in practice of $O(n)$ in both 2D independent and 3D phenomenological models, even with UBS and path compression omitted from the implementation.

\section{\label{sec:results}Results}
\subsection{\label{subsec:saturation}Saturation regimes}
\begin{figure*}
    \centering
    \subfloat[\label{subfig:ds-per-m-naive}]{
        \includegraphics{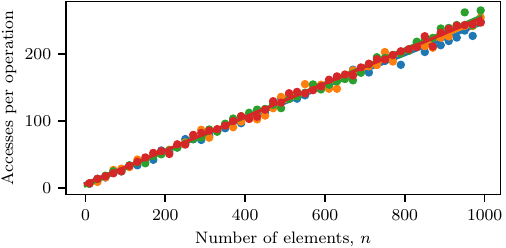}
    }
    \subfloat[\label{subfig:ds-per-m-big-naive}]{
        \includegraphics{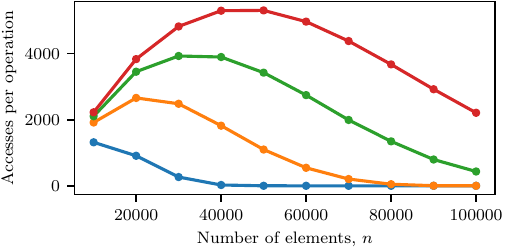}
    }
    
    \subfloat[\label{subfig:ds-per-m-ubs}]{
        \includegraphics{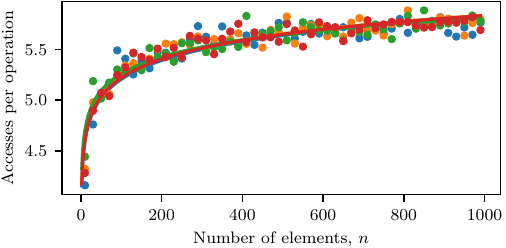}
    }
    \subfloat[\label{subfig:ds-per-m-big-ubs}]{
        \includegraphics{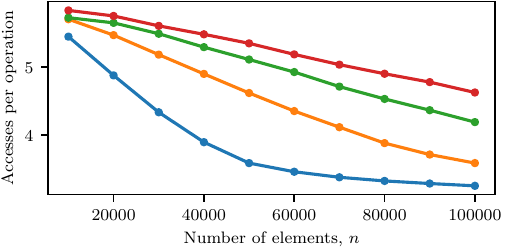}
    }
    
    \subfloat[\label{subfig:ds-per-m-ubs-compression}]{
        \includegraphics{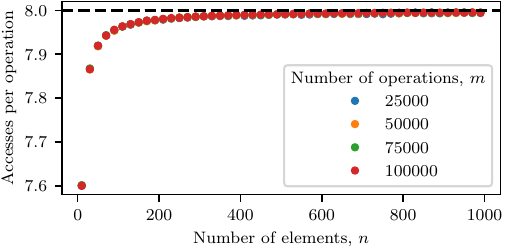}
    }
    \subfloat[\label{subfig:ds-per-m-big-ubs-compression}]{
        \includegraphics{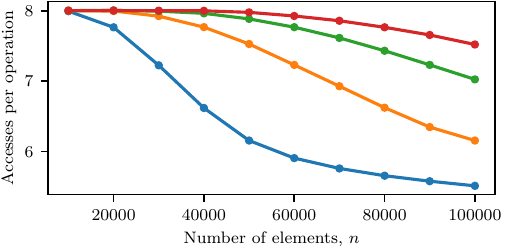}
    }
    \caption{Monte Carlo simulation showing the amortised complexity of operations on the disjoint-set data structure. Each point is the number of root table accesses on a forest of size $n$ taken by $m$ random merge operations, divided by $m$. (a) shows a naive approach, (c) shows UBS and (e) shows both UBS and path compression, demonstrating linear, logarithmic and near-constant scaling, respectively. (b), (d) and (f) extend these plots to higher $n$, showing a change of the scaling in the unsaturated regime.}
    \label{fig:ds-per-m}
\end{figure*}

Fig.~\ref{fig:ds-per-m} demonstrates the amortised time complexity of operations on the disjoint-set data structure via a Monte Carlo simulation under various modes of operation, constituting $m$ merge operations between two random elements in a forest of size $n$. For $n \ll m$, a linear growth in the number of root table accesses is observed with $n$ under naive implementation. This changes to logarithmic with union-by-size (UBS) and near-constant with both UBS and path compression. Specifically, in the latter case, the number of accesses converges upon a local (but effectively global) constant of exactly 8, because asymptotically effectively all elements become a direct descendant of a root, yielding exactly 4 accesses per find operation. Therefore, we denote this mode of operation the \emph{saturated regime}.

With $n \ll m$, the emergent behaviour accurately reflects the theoretical amortised complexity because the number of operations $m$ is effectively infinite. However, if $n \centernot\ll m$, the characteristic trees do not have opportunity to form, such that the complexity is seen to degenerate -- the heights of the trees are now bounded by $m$, not $n$. As the trees of characteristic depth do not dominate the forest, we denote this mode of operation the \emph{unsaturated regime}.

\subsection{\label{subsec:redundancy}Redundancy of optimisations}
Initially, basic timing experiments were performed where the real time taken by the AFS simulation to decode 5000 random instances was measured. The naive approach and UBS performed identically, whereas adding path compression caused the time taken to increase, especially prevalent at greater $p$ and $d$.

\begin{figure}
    \centering
    \subfloat[\label{subfig:scale-factor-ubs}]{
        \includegraphics{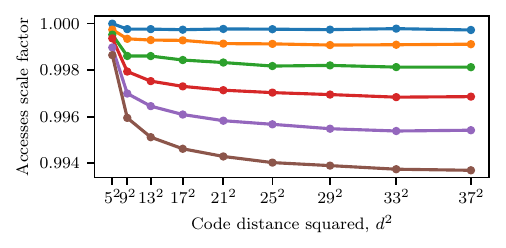}
    }
    
    \subfloat[\label{subfig:scale-factor-compression}]{
        \includegraphics{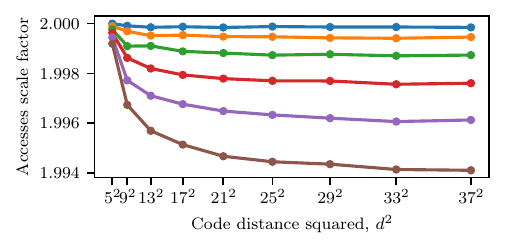}
    }
    
    \subfloat[\label{subfig:scale-factor-ubs-compression}]{
        \includegraphics{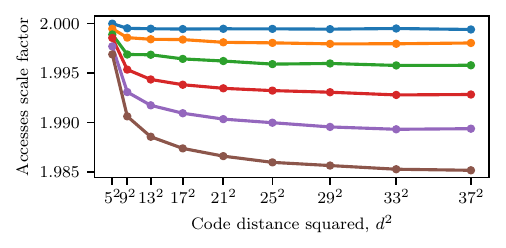}
    }
    
    \subfloat{
        \includegraphics{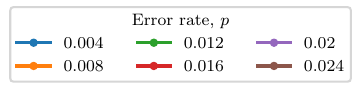}
    }
    \caption{Monte Carlo simulation showing the scale factor in the number of root table accesses when applying (a) UBS, (b) path compression and (c) both UBS and path compression to the graph generator. Each point is the total number of reads over $10^5$ runs, divided by the corresponding number from the naive implementation.}
    \label{fig:scale-factor}
\end{figure}

Fig.~\ref{fig:scale-factor} shows the scale factor in the number of root table accesses when adding UBS and path compression; that is, the number of accesses divided by the corresponding number from the naive implementation. The number of accesses decreases marginally when adding UBS, although this then requires the maintaining and accessing of a size table in conjunction. In the worst case, adding path compression (or path splitting) exactly doubles the number of accesses, decreasing only marginally below this at scale.

For example, a simulation with $d=49$ and $p=0.08$ required $4818.79$ root table accesses on average (mean over $10^6$ instances) with a naive implementation. With UBS, this was reduced to $4489.41$, but also required $3722.97$ size table accesses, thus requiring $3393.59$ more accesses overall. When adding path compression, the number of size table accesses remained the same, but the number of root table accesses increased to $7841.13$, thus requiring $11564.1$ more accesses than the naive implementation overall.

It is apparent that both UBS and path compression not only fail to provide an improvement in decoding time, but in fact tend to worsen it at scale. In the case of UBS, marginal reduction in root table accesses is dwarfed by the additional overhead of maintaining and accessing a size table. In the case of path compression, root table accesses are strictly increased, with the additional overhead of tree traversal registers. Even if path splitting is used in place of compression -- saving the overhead of registers and a second pass -- the increase in the number of root table reads in identical.

The reasoning for this is twofold. First, it is apparent that the data structure is operating in the unsaturated regime. Recall that this means that the complexity degenerates if the sets are bound by their population rate, not by the size of the structure. In percolation theory, if the population rate of clusters is below some threshold, the clusters sizes are bound solely by population, not by the size of the lattice \cite{staufferIntroductionPercolationTheory2018}. We use this analytical model to prove that the data structure operates in the unsaturated regime below in Section~\ref{subsec:percolation}.

Second, the complexity analysis in Section~\ref{subsec:disjoint-set} assumes the use of merge operations between two elements in the forest selected randomly and uniformly. In the context of the decoder, however, merge arguments are not uniformly distributed and are instead much more likely to be at least one cluster root. This is due to the fact that the decoder iterates over odd and unconfined cluster to grow and therefore begins growth with preexisting knowledge of the root. In the simplest yet ubiquitous case, a cluster merges with an unpopulated vertex -- with both arguments roots of their respective sets, no traversal at all is required. This nonuniform operation acts to naturally limit the tree heights formed without requiring explicit path compression.

\subsection{\label{subsec:percolation}Erasure percolation}
In percolation theory, clusters are formed on a lattice of size $L$ by population of neighbouring sites (or bonds between sites) with independent rate $p$. If a cluster spans between two opposite lattice boundaries, it is said to \emph{percolate}. A key observation is that there asymptotically exists a state transition with a percolation threshold $p_c$, below which is a sparse regime with no percolating clusters, and above which exists a single, percolating cluster. In the sparse regime where $p \ll p_c$, the average cluster size scales only with $p$, not $L$, for $L \to \infty$ \cite{staufferIntroductionPercolationTheory2018}.

\begin{figure}
    \centering
    \includegraphics{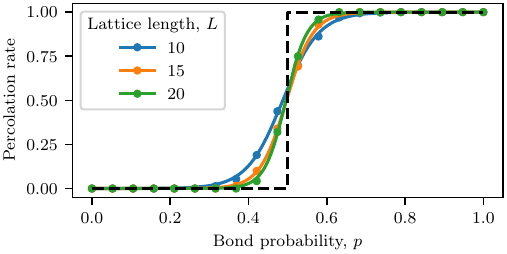}
    \caption{Monte Carlo simulation of bond percolation on a 2D square lattice, demonstrating logistic fits converging to a step function at $p_c=0.5$ for $L \to \infty$. Each point is obtained from 500 samples.}
    \label{fig:square-percolation}
\end{figure}

Fig.~\ref{fig:square-percolation} demonstrates the emergence of characteristic sigmoid curves when plotting percolation rate as a function of $p$, which converge to a step function at the threshold $p_c$, which is exactly 0.5 for bond percolation on a 2D square lattice.

\begin{figure}
    \centering
    \subfloat[\label{subfig:mean-cluster-size}]{
        \includegraphics{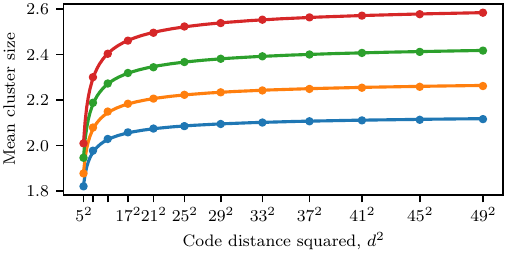}
    }
    
    \subfloat[\label{subfig:mean-cluster-perimeter}]{
        \includegraphics{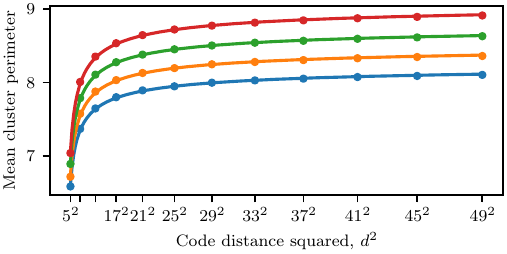}
    }
    
    \subfloat[\label{subfig:mean-cluster-number}]{
        \includegraphics{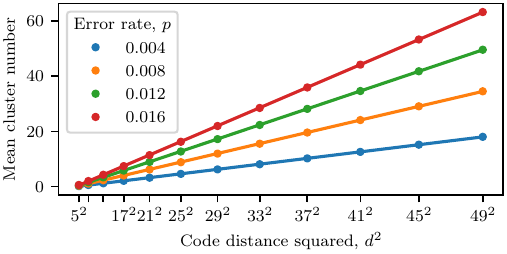}
    }
    
    \caption{Sub-threshold, the mean cluster size (a) and perimeter (b) is bound solely by $p$, when accounting for hyperbolically-decaying boundary effects. The mean number of clusters (c) scales linearly with $d^2$. Each point is obtained from $10^5$ runs.}
    \label{fig:mean-cluster}
\end{figure}

Consider the erasure clusters formed by syndrome validation. Boundary effects at small $d$ limit cluster sizes, preventing a truly constant plot with $d$. The prevalence of clusters incident to the boundary tends to zero with $1/d$. Thus, the mean cluster size follows a plot of $A-B/d$, where $A$ is dependent on the error rate $p$ (Fig.~\ref{fig:mean-cluster}). We also define the \emph{perimeter} of a cluster, which can be easily defined as the length of the list of boundary sites -- this follows similar scaling to the size. The number of clusters, however, scales linearly with $d^2$ ($d^3$ in 3D) assuming a uniform error distribution.

The time taken in each find operation depends on the height of the trees formed in the forest. In the sparse regime, the tree heights are -- as with cluster size -- invariant with $d$ barring hyperbolically-decaying boundary effects.

If it is shown that the erasure clusters exist in a sparse regime (as is implied by the trends in Fig.~\ref{fig:mean-cluster}), then it is clear that the data structure operates in an unsaturated regime, with average cluster sizes depending only on $p$ and invariant with $d$. Clusters are grown from syndrome vertices, which have a nontrivial population distribution arising from $p$. \emph{Syndrome percolation} occurs when there exists a path of neighbouring syndrome vertices between opposite boundaries of the lattice \cite{anwarFastDecodersQudit2014}. However, we are specifically considering the size of erasure trees formed at the conclusion of syndrome validation. We define \emph{erasure percolation} as the existence of an erasure tree spanning between opposite boundaries of the lattice. This is a more relaxed definition than syndrome percolation, as it can arise, for example, from fewer syndrome vertices positioned equidistantly across a dimension, requiring multiple growth iterations.

Therefore, not only is this abstracted from qubit error distributions to syndrome distributions, but to the distribution of all vertices included in grown clusters. In order to model the erasure percolation threshold at higher dimensions without requiring full implementation of syndrome validation, we find a minimum-weight perfect matching (MWPM) on a syndrome graph and estimate the erasure trees by including all vertices within distance $\lfloor w/2 \rfloor$ from each syndrome in a matching, where $w$ is the weight of the matching. 

\begin{figure}
    \centering
    \subfloat[\label{subfig:erasure-percolation-2D}]{
        \includegraphics{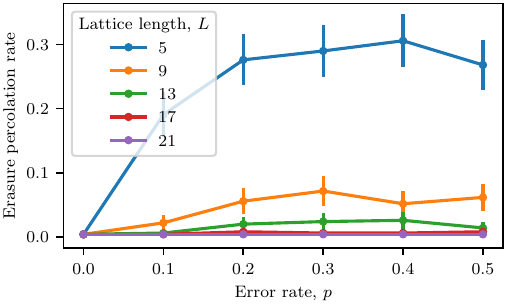}
    }

    \subfloat[\label{subfig:erasure-percolation-3D}]{
        \includegraphics{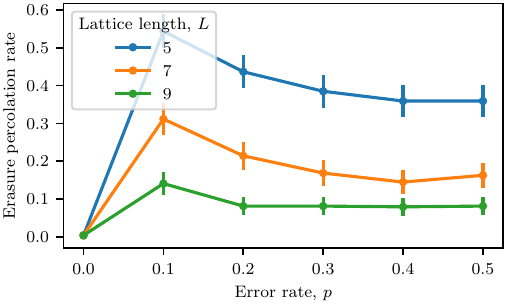}
    }
    
    \caption{Erasure percolation with 2D independent (a) and 3D phenomenological (b) error models ($q=p$). Instead of sigmoids showing a threshold value, percolation rates strictly tend to zero for $L \to\infty$. Each point is obtained from 500 samples using the Wilson score with two standard deviations (${\sim}95\%$ confidence interval) \cite{newcombeTwosidedConfidenceIntervals1998}.}
    \label{fig:erasure-percolation}
\end{figure}

If an erasure percolation threshold exists and is greater than the code threshold, then it follows that the decoder will always operate the data structure in the unsaturated regime. Fig.~\ref{fig:erasure-percolation} shows the results of a Monte Carlo simulation of syndrome validation on toric codes with both 2D independent and 3D phenomenological error models. Usually, one would expect sigmoid plots converging on a step function at $p=p_c$ for $L\to\infty$, as in Fig.~\ref{fig:square-percolation}. Instead, we see that the erasure percolation rate converges to zero for all $p$ for $L\to\infty$. This indicates that an erasure percolation threshold does not exist for any mode of operation of the union--find decoder; that is, the decoder at scale cannot produce syndromes yielding erasure percolation. As the average size of erasure clusters will never be bound by the size of the lattice (ignoring boundary effects on small lattices), the disjoint-set data structure operates strictly in the unsaturated regime, confirming that both union-by-size and path compression are asymptotically irrelevant.


\subsection{Summary}
In Section~\ref{subsec:disjoint-set}, we described how the worst-case complexity of a naive implementation of the union--find decoder (i.e.\ forgoing union-by-size and path compression) would be expected to  be $O(n^2)$, as opposed to $O(n\alpha (n))$ when including the optimisations. However, our numerical data does not support this. The derivation of that scaling assumes that the complexity of each find operation is $O(n)$ in the naive case, which is true only if the size of clusters varies with the lattice size. Our results demonstrate that the nature of the decoder algorithm leads to a strictly sparse (i.e.\ never-percolating) regime of erasure clusters. Therefore, the number of clusters may be linear in $n$, but their average size depends only on $p$, instead yielding an overall complexity of just $O(n)$. This absence of a percolation threshold has been demonstrated in both 2D independent and 3D phenomenological models. In other words, the probability of the worst-case complexity scenario (of erasure clusters percolating) is suppressed to zero for $d\to\infty$.

Therefore, we suggest that both union-by-size and path compression may be comfortably omitted from implementations of the union--find decoder without paying a penalty, and indeed memory usage and runtime are actually improved. In the example of the AFS architecture in \cite{dasAFSAccurateFast2022}, this would involve omitting the size tables and traversal registers, as well as their associated logic. The size table is stated to be the single most memory-consuming component (e.g.\ 54.9 KB out of a total 133 KB for $d=25$), so forgoing union-by-size would yield significant gains in a memory-critical cryogenic environment.

\section{Conclusion}
We have explored the asymptotic behaviour of the disjoint-set data structure and shown that it is significantly underutilised in a union--find decoder at scale, running with time complexity lower than theoretically predicted. Even when omitting the optimisations of union-by-size and path compression, the decoder's worst-case complexity rapidly converges to $O(n)$ for $d\to\infty$. We have argued, therefore, that these optimisations are unnecessary at best and detrimental at worst.

By showing that no percolation threshold exists for the formation of erasure clusters, we have shown that the erasure clusters are strictly sparse in the disjoint-set data structure for $d \to\infty$. Future work could explore the implications of this bound on the decoding strategy, such as the distributed decoding of erasure clusters.

We have also outlined an alternative implementation of syndrome validation for the AFS architecture, which stores boundary lists (instead of recalculating them in each iteration) whilst minimising overhead via double-buffering with a \emph{new edge stack} (NES). Depending on the exact method with which it is contrasted, we suggest this may save computational time over the course of decoding at the slight expense of storage and merging.

\begin{acknowledgments}
This work was supported by University College London, Riverlane and the Engineering and Physical Sciences Research Council [grant number EP/S021582/1]. We'd like to thank James Cruise and Neil Gillespie for helpful and insightful discussions.

\end{acknowledgments}

\bibliography{apssamp}

\appendix
\section{\label{appendix:afs}AFS implementation and improvements}
Whilst \cite{dasAFSAccurateFast2022} describes the practice of recalculating cluster boundaries in each iteration of growth during syndrome validation, our simulations store and update lists of boundary sites, with the benefit of this data being readily available for analysis (as used in Section~\ref{subsec:percolation}).

In the AFS architecture, a \emph{fusion edge stack} (FES) is used to prevent double-growth by delaying the writing of edges between newly-connected clusters to the STM. As we are now storing boundary lists, we also use a \emph{new edge stack} (NES) to store and delay the writing of \emph{all} newly-grown edges, to prevent a similar issue. A \emph{confinement register} (one bit per vertex) is also used to disable the growth of clusters which meet the open boundary.

A notable benefit of this approach is that the NES, by definition, contains exactly the new boundary list for a grown cluster. Therefore, the boundary list and NES can operate as a double buffer, wherein the NES becomes the new boundary list. Swapping the two lists and clearing the new NES (now containing the old boundary data) ready for reuse can both be performed in $O(1)$ (constant) time.

The main drawback of this approach is that boundary lists must be concatenated when clusters are to be merged. Using arrays, this has time complexity linear in the size of the shorter list. Using linked lists, this can be improved to constant time, but this is unlikely to be worthwhile overall due to cache inefficiency and roughly doubling memory usage.

\end{document}